\def\del{\partial}
\def\t{\theta}
\def\<{\langle}
\def\>{\rangle}

\documentstyle[twocolumn,prl,aps]{revtex}
\begin{document}
\twocolumn[\hsize\textwidth\columnwidth\hsize\csname
@twocolumnfalse\endcsname

\title{Exact non-equilibrium DC shot noise in Luttinger liquids and
\\ fractional quantum Hall devices }
\author{P. Fendley$^1$, A.W.W. Ludwig$^{2\dagger}$ and H.
Saleur$^{1*}$}
\address{$^1$ Department of Physics, University of Southern
California,
Los Angeles CA 90089-0484}
\address{$^2$ Physics Department, University of California,
Santa Barbara, CA 93106}
\date{May 1995, cond-mat/9505031}
\maketitle
\begin{abstract}
A point contact in a Luttinger liquid couples
the left- and right-moving channels, producing shot noise.
We calculate exactly the DC shot noise at zero temperature
in the out-of-equilibrium  steady state where current is flowing.
Integrability of the interaction
ensures the existence of a quasiparticle
basis where quasiparticles scatter ``one by one'' off the
point contact. This
enables us to apply a direct generalization of
the Landauer approach to shot noise to this interacting model.
We find a simple relation of the
noise to the current and the differential conductance.
Our results should be
experimentally-testable in
a fractional quantum Hall effect device,
providing a clear signal of the fractional
charge of the Laughlin quasiparticles.
\end{abstract}
\pacs{PACS numbers:  ???}

]

\narrowtext

The Luttinger model, describing the low-energy excitations
of an interacting one-dimensional fermion gas,
is one of the simplest non-Fermi-liquid metals.
Experimental observation of this non-Fermi state
in 1D quantum wires is difficult
since disorder tends to localize these excitations.
However, this theory has been proposed to describe
the edge states in fractional quantum Hall effect devices
\cite{Wen}. Tunneling through
a point contact in a Luttinger liquid is a
practically ideal situation for making contact
between experiment and theory \cite{KF,Moon,FLSi}. Very clean
measurements of  electronic transport properties
through point contacts in quantum Hall devices have
been performed \cite{exper}, while there are
powerful constraints on the theory because the
model is integrable \cite{GZ}.
One can compute the current and conductance through
the point contact exactly, even when the system
is out of equilibrium \cite{FLSi,FLSii,FLSjack}.
The  experimentally-measured \cite{exper}\
linear-response conductance in such a device
 agrees well with the  exact theoretical prediction
\cite{KF,Moon,FLSi}.

The tunneling current and conductance
are ``spectroscopic'' probes of
the non-Fermi-liquid state in the leads.
These however are not the only transport
properties of interest.
The current shot noise resulting from the point contact
provides another signal of the non-Fermi-liquid state.
Here we compute the
zero-temperature DC shot noise exactly.
This is the first exact computation of noise in a model
with interacting electrons.

For weak backscattering, Laughlin quasiparticles
hop from one edge to the other at the point contact.
For strong backscattering, on the other hand,
current transport is caused by electrons.
In both limits the tunneling events happen
 independently, so
the shot  noise is
proportional to the charge of the carriers providing
transport. The weak-backscattering limit is thus a direct
signal of the fractional charge of the Laughlin quasiparticles
in the Hall device \cite{KFnoise}.
Our non-perturbative results give the noise for any amount
of backscattering. Moreover,  we find a simple expression
of the noise in terms of the current and conductance.

The shot noise
is a function of the (driving) ``source-drain'' voltage
$V$, and $T_B$,
the scaled point contact interaction strength.
Our approach uses the exact bulk and impurity $S$
matrices \cite{ZZ,GZ}\ of the Luttinger model.
The bulk $S$ matrix describes the scattering of
quasiparticles off each other away from the point contact.
Knowing it allows us to find the density of these quasiparticles in
the Fermi sea at any voltage.
The impurity $S$ matrix elements give the probability
for tunneling events, which correspond to
the scattering  of the quasiparticles
off the point contact.
We find the zero-temperature
DC noise using a Landauer-type approach
familiar from studies of transport of free electrons
 \cite{Lesovik,Buttiker,Landauer}; the result
involves the
tunneling probabilities and the quasiparticle density.
It is exact
in this integrable system
because the quasiparticles are transported
``one by one''.
This means that the scattering of a current of quasiparticles
off the point contact can be described by the product of
one-particle $S$ matrices, even though the quasiparticles interact.

We first review the bosonized formulation of the
Luttinger model, before displaying the appropriate
quasiparticle basis.
The left- and right-moving channels of the Luttinger model
are described by left- and right-moving bosons
$\phi_L$ and $\phi_R$ defined on a space $-l < x < l$ \cite{Lutt}.
The coupling constant of the fermion model
is parametrized by $\nu$, which is the filling fraction
of the Hall device when $1/\nu$ is an odd integer.
In the absence of the
point contact, the Hamiltonian is of Tomonaga form:
\begin{equation}
H_0={ v_F \pi\over \nu} \int_{-l}^l dx
\left[j_L^2+j_R^2\right],
\label{hamil}
\end{equation}
quadratic in the two individually-conserved $U(1)$ currents,
$j_L=-{1\over 4\pi}(\partial_t+\partial_x)\phi_L$ and $j_R={1\over
4\pi} (\partial_t-\partial_x)\phi_R$. These currents are the charge
densities: for example, $ e Q_L= e\int_{-l}^ldx j_L$
is the total charge of the left-moving channel.
An applied  voltage $V$ imposes  a chemical
potential difference
for the injected  left- and
right-moving charge carriers. This results in
a term $-\Delta Q eV/2 $ in the Hamiltonian, where $\Delta Q
\equiv (Q_L - Q_R)$.

A point-contact interaction  coupling
the right and left channels at the origin $x=0$
results in backscattering, which in the FQHE edge state model
corresponds to
tunneling of Laughlin quasiparticles.
The resulting Hamiltonian includes the term \cite{KF}
\begin{equation}
H_B =\lambda \cos\left[\phi_L(x=0) - \phi_R(x=0)\right] .
\label{delH}
\end{equation}
Other allowed terms are
irrelevant for $\nu>1/4$ \cite{KF}; our analysis  holds for  any
$\nu$
as long as there is only a single relevant operator
 in $H_B$.
We rewrite the model in terms of two left-moving bosons \cite{WA}:
\begin{equation}
\phi^{e,o}(x+t) \equiv
{1 \over \sqrt{2}} [ \phi_L(x,t) \pm\phi_R(-x,t)],
\qquad
\label{evenodd}
\end{equation}
where the {\it even} boson $\phi^e$ and the {\it odd} boson
$\phi^o$ have the $+$ and $-$ sign, respectively.
The even and odd Hamiltonians are decoupled:
$H_B$
involves only the odd boson, while
in $H_0$ $j_L,j_R$ are replaced with $j^e, j^o$,
where $j^{e,o}(x+t)= (1/\sqrt{2}) [ j_L(x,t) \pm
j_R(-x,t)]$.
The even
and odd charges are thus related to the charges of the original left-
and right-moving edges by $\Delta Q = Q_L-Q_R =\sqrt{2} Q^o $ and
$Q_L+Q_R = \sqrt{2} Q^e$. Therefore $Q^e$ is the total charge on
both edges and is conserved even in the presence of the interaction.
The backscattering current thus depends only on the odd boson theory.

Describing the model in terms of quasiparticles allows us to
calculate exact transport properties.
These quasiparticles span the Hilbert space
of the left-moving odd boson; they are the excitations above
the ``Fermi sea'' (at zero voltage).
Because
the odd boson theory is integrable \cite{GZ},
these quasiparticles have very
special properties. The
infinite number of conserved quantities which commute with
the Hamiltonian in an integrable model
define a basis of quasiparticles
where the quasiparticles scatter ``one-by-one''.
The conservation laws allow determination of
the exact quasiparticle spectrum and the $S$ matrix.
This result is already known \cite{ZZ,GZ,FSW}\ for
the odd boson Hamiltonian (\ref{hamil}) and (\ref{delH}).
For any $\nu$, the spectrum contains a kink ($+$) and an
antikink ($-$). These carry
(odd) charges $ Q^o =
1/\sqrt{2}$ and $-1/\sqrt{2}$,
respectively.
At some values of $\nu$,  there are chargeless ``breather'' states,
but these do not enter into the zero-temperature analysis below.
We parametrize the energy and momentum
of these massless left-moving quasiparticles in terms of the
rapidity $\theta$ defined by $E= -pv_F= Me^\t/2$, where $M$ is
an arbitrary energy which cancels out of physical quantities.

When a positive
voltage is turned on,
it becomes energetically favorable for positively-charged
particles (the kinks) to fill the sea. If the quasiparticles did not
interact, all momentum states with $v_Fp < eV/2$
 would be filled with kinks
(no antikinks) at zero temperature.
The interaction affects not only the  position of the
Fermi level but also
the density of quasiparticles. In \cite{FLSi,FLSii} it was
shown
how to find the shift of the Fermi level due to an applied $V$
exactly, following
techniques given for example in \cite{YY}.
We define the density $ \rho(\t)$ so that
$\rho(\theta)d\theta$ is
the number of  {\it kinks} per unit length with rapidities between
$\theta$ and $\theta+d\theta$. The shift of the Fermi level
for kinks is given by the
quantity $A$, such that $\rho(\theta)=0$ for $\theta >A$.
Then
\begin{eqnarray}
\tilde{\rho}(s)&=&{M\over 2i hv_F}
{K(-s)K(i)\over s-i}e^{(is+1)A}.\\
\label{forrho}
e^A&=&{eV\over M}{K(0)\over K(i)}
\end{eqnarray}
where $\tilde\rho(s)$ is the Fourier transform of $\rho(\t)$, and
\begin{eqnarray}
\nonumber
K(s)&\equiv&\sqrt{2\pi\over \nu}
{\Gamma\left(s/[2i(1-\nu)]\right)\over
\Gamma\left(\nu s/[2i(1-\nu)]\right)
\Gamma\left(1/2-is/2\right)} e^{-is\Delta}\\
\nonumber
\Delta&\equiv&{\nu\over 2(1-\nu)}\ln \nu-{1\over 2}\ln(1-\nu)
\end{eqnarray}
As required, $\rho(\theta)=0$ for $\theta >A$,
because $K$ is analytic everywhere in the upper half plane
(including $s=i\infty$).

Without the backscattering interaction, the odd charge is conserved
($\partial_t \Delta Q =0$). Because there are no
antikinks at zero temperature,
all the current arises from the kinks moving to the left at the
fermi velocity:
\begin{equation}
I_0(V)= ev_F \int_{-\infty}^A
d\t \rho(\t) = ev_F \tilde\rho(0)= \nu{e^2\over h}V
\label{izero}
\end{equation}
The backscattering current $I_B(V)$ is the rate
at which the charge of the left-moving edge is depleted due
to backscattering off the impurity. This
decreases the total current $I=I_0+ I_B$.
By symmetry, $\partial_t Q_L = - \partial_t Q_R$,
so $I_B =  \partial_t  \bigl ( {e\over 2}\Delta Q \bigr ) =
\partial_t  \bigl ( {e\over \sqrt{2}} Q^o \bigr )$.
In the even/odd basis, tunneling
corresponds to the violation of odd charge conservation
at the contact.
In the quasiparticle basis, this corresponds (at $T=0$)
 to a kink
scattering off the contact into an antikink.

At $T=0$, both the backscattering current and the DC shot noise
can be written  entirely in terms of the tunneling probability
and the density
$\rho$  of {\it kinks} only.
The impurity $S$ matrix element
$S_{jk}(p/T_B)$ describes
a single quasiparticle of type $j$ and momentum $p$ scattering
elastically off the point
contact into a quasiparticle of type $k$.
Here $T_B \propto \lambda^{1/(1-\nu)}$ is
the crossover scale introduced
by the interaction. We define
$T_B\equiv Me^{\theta_B}/2$ so that the $S$ matrix
elements are functions of $\t-\t_B$.
These were derived exactly in \cite{GZ};
the tunneling probability is given by
\begin{equation}
|S_{+-}(\t-\t_B)|^2={1
\over 1+ \exp[2(1-\nu)(\t-\t_B)/\nu]}
\label{bdrs}
\end{equation}
and $|S_{++}|^2 = 1- |S_{+-}|^2$.
A simple kinetic equation then gives \cite{FLSi,FLSii},
when specialized to $T=0$
\begin{equation}
I_B(V,T_B)= -ev_F \int_{-\infty}^A
d\t\ \rho(\t) |S_{+-}(\t-\t_B)|^2
\label{iback}
\end{equation}
The differential conductance is $G=\del_V I$.
Using the explicit expressions for $\rho$ and $|S_{+-}|^2$, one
finds power series expressions for
$I(V,T_B)=I_0(V)+I_B(V,T_B)$; they are
\begin{eqnarray}
\label{powercurrent}
I(V,T_B)&=& {e^2V\over h}\left[\nu-\nu^2\sum_{n=1}^\infty
 a_n(\nu) \left({eV\over T_B'}\right)^{2n({\nu}-1)}\right]\\
\label{powercurrentii}
I(V,T_B)&=& {e^2V\over h} \sum_{n=1}^\infty a_n(1/\nu)
\left({eV\over T_B'}\right)^{2n({1\over \nu}-1)}
\end{eqnarray}
where
\begin{eqnarray}
\nonumber
a_n(\nu)&=&(-1)^{n+1}
{\sqrt{\pi}\,
\Gamma({n\over \nu})\over 2\Gamma(n)\Gamma({3\over 2}+n({1\over
\nu}-1))}\\
\nonumber
T_B'&\equiv& T_B {2\sqrt\pi \Gamma(1/[2(1-\nu)]) \over
\Gamma(\nu /[2(1-\nu)])}
\end{eqnarray}
The expansion (\ref{powercurrent}) is appropriate for $T_B/V <1 $,
while (\ref{powercurrentii}) is appropriate for $T_B/V >1$;
notice the strong barrier/weak barrier duality \cite{FLSii}.

Finding the shot noise requires a more detailed analysis.
We follow arguments  given in \cite{Landauer}\ for the free-electron
case, and show that they can
be directly generalized
to our interacting quasiparticles,
due to the constraints of integrability.
To make the calculation of the noise precise, we first examine
the quantum-mechanical current operator $j(t)$, which includes
the current with its fluctuations. The system is not in equilibrium
because current is flowing, but it is in a steady state, so
the current $I=\<j(t)\>$ does not
depend on time. The
current fluctuations in frequency space are characterized
by the correlator
\begin{equation}
C(\omega)= {1\over 2} \int dt\, e^{i\omega t}
\<\{j(t),j(0)\}\>.$$
\label{noisecorr}
\end{equation}
We will focus  on the noise at zero frequency (the DC limit),
which we denote by $\<I^2\>\equiv C(0)$.

In our quasiparticle approach, the current is thought of
as a series of individual quasiparticles. Since the model is interacting,
the quasiparticles are correlated, but at zero temperature
every kink state with rapidity (parametrizing momentum)
 less than $A$ is filled, and
the remaining kink states as well as all antikink states
are empty. Thus without
the point contact there is no noise in this steady state.
(Even with the impurity there is no noise in the even current,
only in the odd current.)
When the backscattering is included,
shot noise occurs because there are two possible outcomes
when a given quasiparticle hits the impurity.
We can describe the DC shot noise
from a quasiparticle approach, because the scattering off
the point contact is elastic and one-by-one.
As we discussed above,
when a left mover backscatters into a right mover, in the even/odd
basis this corresponds to an odd-boson kink scattering into an
antikink.
Thus as we turn on the interaction (and the
voltage in order to generate a  population of
kinks) the impurity will scatter
some of these into antikinks. The bath of kinks (the battery)
is large so it is not depleted by the scattering.

Consider a single kink of momentum $p$, and define $f=1$
if this kink turns into an antikink when it scatters off the
impurity, and $f=0$ if it scatters into a kink.
By definition of the impurity $S$ matrix elements,
the average over many events is
$\<f\>=|S_{+-}(p/T_B)|^2$.
In the quasiparticle approach, the noise is then
proportional to the  the fluctuation of $f$
$$ \<I^2\> =  e^2 \sum \<(f-\bar
f)^2\>=e^2\sum\left(\<f^2\>-\<f\>^2\right)
$$
where the sum is over all kink states which hit the point
contact per unit time.
The crucial point  to notice is that because $f$ is either $0$ or
$1$,
$\<f^2\>=\< f\>$ \cite{Landauer}. We replace the sum with
an integral over momenta, obtaining
\begin{eqnarray}
\nonumber
\<I^2\> = e^2 v_F \int_{-\infty}^A  d\t\ \rho (\theta)
&&|S_{+-}(\t-\t_B)|^2 \times \\
&& (1-|S_{+-}(\t-\t_B)|^2)
\label{noislesov}
\end{eqnarray}
This form of the DC shot noise is formally
the same as for non-interacting
electrons
\cite{Lesovik,Buttiker,Landauer}; but here the
density of states $\rho(\t)$ given by
(\ref{forrho}) is non-trivial.

Since we have explicit expressions for $|S_{+-}|^2$
 and $\rho$,
power series expressions for the noise analogous to
(\ref{powercurrent}) and
(\ref{powercurrentii}) can be found directly from
(\ref{noislesov}). This however is not necessary
because this expression can be related directly  to the current.
The specific form (\ref{bdrs}) of the transmission amplitude
means that we can write
\begin{equation}
|S_{+-}|^2(1-|S_{+-}|^2)= {\nu\over 2(1-\nu)}
{\del |S_{+-}|^2 \over \del \t_B}.$$
\label{identS}
\end{equation}
Since neither $\rho$ nor $A$ depends on $\t_B$,
we can pull the $\del_{\t_B}$ out of the integral.
Using the expressions (\ref{izero}) and
(\ref{iback}) for $I(V,T_B)$ yields
\begin{equation}
\<I^2\> = -{\nu e\over 2(1-\nu)} T_B \del_{T_B} I(V,T_B)
\label{flucdissi}
\end{equation}
This expression  is a very simple analog
of the fluctuation-dissipation theorem
for a Luttinger liquid at zero temperature.
Since $I(V,T_B)/V$ is a function of only $V/T_B$, we
find another form of this relation:
\begin{equation}
\<I^2\> = {\nu e\over 2(1-\nu)} V^2\del_{V} {I\over V}=
{\nu e\over 2(1-\nu)} (VG-I).$$
\label{flucdiss}
\end{equation}
It is certainly conceivable that (\ref{flucdissi})
and (\ref{flucdiss}) hold for other models, because they are
a consequence of the simple identity (\ref{identS}).
The noise is plotted in figure 1. Remarkably, due to the simple form
of (\ref{flucdiss}), the extrema of $\<I^2\>/T_B$
and $G$ occur at the same value of $T_B/V$.

We can compare these exact equations with known results
in the limits of weak ($|S_{+-}|^2$ small) and strong ($|S_{+-}|^2$
near 1) backscattering. In the weak backscattering limit,
the events happen seldomly and thus should be uncorrelated.
Thus the shot noise should obey the formula for non-interacting
particles (see e.g. \cite{Lesovik,Buttiker,Landauer})
$$\<I^2\>\approx \nu e (I_0-I) \qquad T_B \hbox{ small};$$
the $\nu$ appears because the original Luttinger fermions
(the Laughlin quasiparticles in the FQHE edge realization)
being scattered have charge $\nu e$. Inserting
the small-$T_B$ expansion from (\ref{powercurrent}) into
(\ref{flucdissi}), one
easily verifies this.
Measuring the noise therefore gives a direct measurement of the
fractional charge \cite{KFnoise}. In
the strong backscattering limit, one can check that the
leading irrelevant operator contributing to the current
corresponds to the tunneling of quasiparticles of charge $e$.
In the FQHE this is the tunneling of physical electrons
between two separate systems (in the strong backscattering
 limit the point contact splits the system in two). Thus
$$\<I^2\>\approx e I \qquad T_B \hbox{ large},$$
which is easily verified using the large-$T_B$
expansion (\ref{powercurrentii}).

The interactions between the quasiparticles prevent any
naive application of this formalism outside zero
temperature. However, observe that
if  either $V=0$ or $T_B=0$, the noise
is given by the familiar Johnson-Nyqvist formula $\<I^2\>= 2GT$.
Since this vanishes at $T=0$ and our zero-temperature result
vanishes when either $V=0$ or $T_B=0$, a good approximation
to the noise for arbitrary values of $V,T_B$ and $T$ should be
given by adding the two types of contributions:
\begin{equation}
\<I^2\>(T)\approx{\nu e\over 2(1-\nu)} (VG-I) + 2GT.
\label{noiseguess}
\end{equation}
It is conceivable that such a  formula, with
added terms of the form $T G^m (G-I/V)^n$ ($m,n\ne 0$),
might be exact, generalizing the above
analog of the fluctuation dissipation theorem.

We have seen that integrability
permits  the  exact calculation of
  non-equilibrium transport properties
through (interacting) point contacts
in a Luttinger liquid, generalizing notions of ballistic transport
used previously only for non-interacting electrons.
It would be very interesting to compare
our exact findings with future experiments
on shot noise in Quantum Hall devices, identifying
the fractional charge of the quasiparticles
experimentally.
It would also be most interesting
to extend these results to non-zero frequency $\omega$ in order to
check the recent perturbative result that the noise has a singularity
in $\omega$ \cite{CFW}. This, however,  might require a more
complicated
formalism based on form factors.

\noindent $^{\dagger}$  A. P. Sloan Fellow.
\noindent $^*$ Packard Fellow.

We thank M.P.A.\ Fisher for useful discussions.
This work was supported by the Packard Foundation,
the A.P.\ Sloan Foundation,  the
National Young Investigator program (NSF-PHY-9357207) and
the DOE (DE-FG03-84ER40168).

\bigskip\bigskip\bigskip\bigskip

\noindent Figure 1.  The exact DC noise $\<I^2\>/V$ as a
function of backscattering and driving voltage $T_B/V$.

\end{document}